\documentclass[pra,twocolumn,floatfix,superscriptaddress,amsmath,citeautoscript,aps,longbibliography]{revtex4-2}
\pdfoutput=1
\usepackage[T1]{fontenc}\usepackage[latin1]{inputenc}
\usepackage{dcolumn,graphicx,color,booktabs,microtype,afterpage}
\usepackage[charter,greekuppercase=italicized]{mathdesign}
\usepackage{soul}
\renewcommand{\figurename}{Fig.}
\renewcommand{\tablename}{Table}
\makeatletter\renewcommand{\fnum@figure}[1]{\figurename~\thefigure.}\makeatother
\makeatletter\renewcommand{\fnum@table}[1]{\tablename~\thetable.}\makeatother
\newcount\hh \newcount\mm
\hh=\time \divide\hh by 60
\mm=\hh \multiply\mm by 60 \mm=-\mm
\advance\mm by \time
\def\now{\number\hh:\ifnum\mm<10{}0\fi\number\mm}
\usepackage[colorlinks,plainpages=false,linkcolor=blue,urlcolor=blue,citecolor=blue,pdfpagemode=UseNone,pdfstartview=FitBH]{hyperref}
\newcommand{\bro}{Cu$_4$SO$_4$(OH)$_6$}

\begin{document}

\title{Helical Spin Dynamics in Commensurate Magnets: a Study on Brochantite, Cu$_4$SO$_4$(OH)$_6$}

\author{S.~E.~Nikitin}
\thanks{These authors contributed equally to this work}
\affiliation{Quantum Criticality and Dynamics Group, Paul Scherrer Institut, CH-5232 Villigen-PSI, Switzerland}
\author{Tao Xie}
\thanks{These authors contributed equally to this work}
\affiliation{Neutron Scattering Division, Oak Ridge National Laboratory, Oak Ridge, TN 37831, USA}
\author{A.~Gazizulina}
\affiliation{Helmholtz-Zentrum Berlin f\"ur Materialien und Energie, Hahn-Meitner-Platz 1,
Berlin D-14109, Germany}
\affiliation{Institute for Quantum Materials and Technologies, Karlsruhe Institute of Technology, 76021 Karlsruhe, Germany}
\author{B.~Ouladdiaf}
\affiliation{Institut Laue-Langevin, 71 avenue des Martyrs, F-38000 Grenoble, France}
\author{J.~A.~Rodr\'{i}guez Velamaz\'{a}n}
\affiliation{Institut Laue-Langevin, 71 avenue des Martyrs, F-38000 Grenoble, France}
\author{I.~F.~D\'{i}az-Ortega}
\affiliation{Institute for Materials Research, Tohoku University, Sendai, 980-8577, Japan}
\affiliation{Departamento de Qu\'{i}mica y F\'{i}sica-CIESOL, Universidad de Almer\'{i}a, Ctra. Sacramento s/n, Almer\'{i}a, Spain}
\author{H.~Nojiri}
\affiliation{Institute for Materials Research, Tohoku University, Sendai, 980-8577, Japan}
\author{L.~M.~Anovitz}
\affiliation{Chemical Sciences Division, Oak Ridge National Laboratory, Oak Ridge, TN 37831, USA}
\author{A.~M.~dos~Santos}
\affiliation{Neutron Scattering Division, Oak Ridge National Laboratory, Oak Ridge, TN 37831, USA}
\author{O.~Prokhnenko}
\thanks{Corresponding author: prokhnenko@helmholtz-berlin.de}
\affiliation{Helmholtz-Zentrum Berlin f\"ur Materialien und Energie, Hahn-Meitner-Platz 1,
Berlin D-14109, Germany}
\author{A.~Podlesnyak}
\thanks{Corresponding author: podlesnyakaa@ornl.gov}
\affiliation{Neutron Scattering Division, Oak Ridge National Laboratory, Oak Ridge, TN 37831, USA}

\begin{abstract}
We report the direct observation of a commensurate-ordered antiferromagnetic (AFM) state but incommensurate helical spin dynamics in the natural mineral brochantite Cu$_4$SO$_4$(OH)$_6$ through neutron diffraction and neutron spectroscopy measurements. Inelastic neutron scattering measurements reveal magnon-like excitations with considerable dispersion along the c-axis and almost flat branches in other principal directions, indicating the strong one-dimensional character of the magnetic correlations. We experimentally observe the effect of the uniform Dzyaloshinskii-Moriya (DM) interaction, which elevates the degeneracy of the spin-wave modes shifting them in opposite directions in reciprocal space. The system has a commensurate AFM ground state, stabilized by the anisotropic symmetric Heisenberg exchange interactions, and quasi-one-dimensional chiral spin dynamics due to the antisymmetric DM interaction. Employing linear spin-wave theory, we were able to construct an effective Heisenberg Hamiltonian. We quantify both the symmetric exchange parameters and the DM vector components in Cu$_4$SO$_4$(OH)$_6$ and determine the mechanism of the magnetic frustration. Our work provides detailed insights into the complex dynamics of the spin chain in the presence of uniform DM interaction.
\end{abstract}

\maketitle

\section{Introduction}\label{sec1}

Nowadays, the development of next-generation magnon-based information carriers for computing devices~\cite{Kruglyak_2010,Chumak,Barman_2021} boosted interest in fundamental properties of the magnon excitations in noncollinear antiferromagnets, such as spin spirals and skyrmions~\cite{Daniels,Janoschek,Gitgeatpong,Onishi,Chernyshev2016,Santos1,Santos2}.
Magnon, a quantized spin wave, is expected to transmit a Joule heat-free spin current and is considered a core of spin-wave-based information nanotechnology~\cite{Hog_2017,Hirobe,Kruglyak_2010,Chumak}.

Interatomic exchange of a pair of interacting spins is generally described by the exchange matrix that can be decomposed into the interaction $J_{ii}$ and antisymmetric component $\mathbf{D}_{ij}$, known as the Dzyaloshinskii-Moriya (DM) interaction~\cite{Dzyaloshinsky, Moriya}.
Competition between the symmetric and antisymmetric interactions can give rise to chiral magnetic orders and significantly impact the emerging spin dynamics~\cite{Puszkarski,Zakeri,Moon,Wang,Hog_2017}.
Dynamical spin-spiral correlations in AFM one-dimensional (1D) $S$=1/2 systems are of special interest, given that a spin chain is the simplest of the spin models and a promising candidate for exotic magnetic phases induced by frustration and quantum fluctuations~\cite{Chubukov,Bonner,Hikihara,Sudan,Onishi, wu_TLL_2019, nikitin_multiple_2021}.
The spectral properties of such a system are nontrivial because of quantum fluctuations arising from the combination of low spin value with reduced dimensionality, which can be further enhanced by frustrated next-nearest neighbor coupling.

The DM interaction between two spins, $\mathbf{S}_i$ and $\mathbf{S}_j$, reads $\mathcal{H}_{\mathrm{DM}} = -\mathbf{D}_{ij} \cdot ( \mathbf{S}_i \times \mathbf{S}_j )$, where the direction of the DM vector $\mathbf{D}_{ij} = \mathrm{D} \mathbf{n}_{ij}$ is constrained by symmetry~\cite{Moriya}.
Depending on the crystallographic symmetry, there are two fundamentally different patterns for the $\mathbf{D}_{ij}$.
Staggered vectors $\mathbf{D}_{ij}$ pointing to alternating antiparallel directions along a chain, induce a canting of magnetic moments in a classical AFM, resulting in a weak ferromagnetic moment~\cite{Dzyaloshinsky,Moriya}.
Uniform DM interaction is characterized by parallel orientation of the $\mathbf{D}_{ij}$ for all bonds within a chain.
The uniform $\mathbf{D}_{ij}$ pointing along the easy axis tends to stabilize a spiral spin structure and splits the spin-wave modes shifting the dispersion curves in opposite directions in the reciprocal space~\cite{Gangadharaiah, Karimi, Bocquet, Santos1, Santos2}.

Inelastic neutron scattering (INS) is the leading technique for the detailed characterization of the spin-wave spectra across the entire Brillouin zone (BZ).
Hence, the INS can reveal both the symmetric Heisenberg exchange and the antisymmetric DM interaction.
Though, the experimental confirmations of theoretical findings remain challenging because of the difficulty in obtaining suitable model compounds.
To the best of our knowledge, only a few spin-chain compounds that display spiral spin dynamics due to the uniform DM interaction have been experimentally explored~\cite{Halg, Gitgeatpong}.

\begin{figure}[htb]
\includegraphics[width=0.7\linewidth]{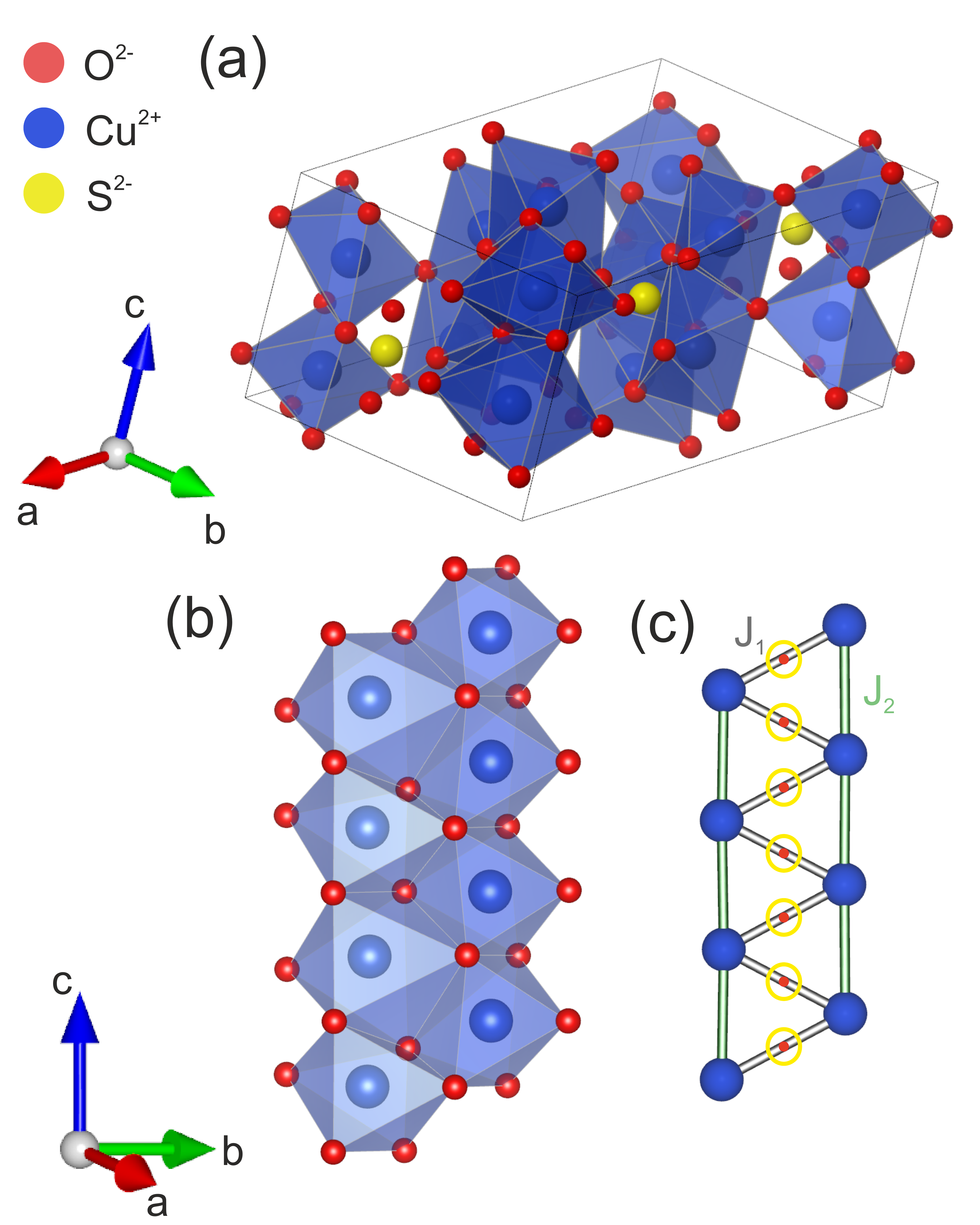}
  \caption{ (a) Crystal structure of brochantite \bro. Red, blue, and yellow balls represent oxygen, copper, and sulfur ions, respectively. Hydroxide anions (OH) are omitted for clarity.  (b) Fragment of the crystal structure, which shows a zigzag Cu chain running along the c axis. (c) The minimal spin model of brochantite is discussed in the main text. Yellow circles at the middle of each $J_1$ bond represent uniform DM interaction.}
  \label{spin-chain}
  \vspace{-12pt}
\end{figure}

Sometimes, nature supplies us with surprise gifts.
Emerald-green crystals of natural brochantite \bro, named after the geologist Andr\'{e} Brochant de Villiers, attract the attention of mineral collectors around the globe. The most common form of brochanite has
monoclinic symmetry P2$_1$/$a$ (Nr. 14) with $a=$13.140, $b =$ 9.863, $c =$ 6.024~\AA, $\beta =$ 103.16$^{\circ}$ as determined from X-ray measurements on natural single crystals~\cite{Merlino}. One has to mention the order-disorder nature of this mineral. In layered materials such as brochantite, neighboring layers can be arranged in many different geometrically and energetically equivalent ways leading to the appearance of either disordered or fully ordered sequences~\cite{Merlino}. The single crystal results agree well with the x-ray and neutron scattering data obtained on hydrothermally prepared synthetic polycrystalline forms~\cite{Vilminot}. This work reported also a magnetic order in brochantite below 7.5~K and proposed an antiferromagnetic structure with the moments aligned along the $a$-axis in the magnetic cell of the same size as the nuclear one.
The scientific interest in brochantite stems from the presence of geometrically frustrated Cu$^{2+}$ ($S$=1/2) magnetic chains (Fig.~\ref{spin-chain}), which turn such a system into a low-dimensional "quantum" magnet that can exhibit emergent phenomena and promotes exotic magnetic states at low temperatures.
It is worth mentioning that synthetic Zn-brochantite, ZnCu$_3$SO$_4$(OH)$_6$, features well-isolated 2D kagome planes and exhibits emergent behavior of quantum spin liquids~\cite{Li2014, Gomilsek2016, Gomilsek2017}.
Note that the synthetic brochantite was obtained as a finely ground powder only~\cite{Vilminot}.
Attempts to promote crystal growth were unsuccessful.

Here, we use thermodynamic measurements, neutron diffraction, INS, and linear spin-wave theory (LSWT) calculations to show that the brochantite is a rare realization of a zigzag spin-chain system with a delicate balance between the symmetric and antisymmetric exchange interactions.
Below the N\'{e}el temperature brochantite exhibits quasi-one-dimensional chiral spin dynamics due to the uniform DM interaction while having the commensurate AFM long-range ground state stabilized by anisotropic Heisenberg exchange.
Our data and analysis of the magnetic excitation modes in brochantite could be a guide for further experimental work and theoretical developments on low-dimensional systems with competing symmetric and antisymmetric interactions.

\section{Experimental details}

The high-quality single crystals of \bro\ were obtained commercially.
Refinement of single-crystal x-ray diffraction data, obtained using Bruker-D8 Advance diffractometer at Helmholtz-Zentrum Berlin (HZB), demonstrated a single brochantite phase.
Magnetization and specific heat measurements were performed at the CoreLab Quantum Materials using Physical Properties Measurement System (PPMS) (Quantum Design) at HZB. High-field magnetization measurements were performed using a 30~T pulsed magnet and a $^4$He bath cryostat at the Institute for Materials Research, Tohoku University (Sendai)~\cite{Nojiri}.

For the characterization of the crystal and magnetic structures, we measured neutron diffraction at the Institute Laue Langevin in Grenoble (France). We used the neutron Laue single-crystal diffractometer Cyclops and two neutron four-circle diffractometers, D9 and D10~\cite{ILLdoi1, ILLdoi2}.
The crystal structure was refined using FullProf software package \cite{FullProf}.

INS measurements were performed at the time-of-flight Cold Neutron Chopper Spectrometer (CNCS)~\cite{CNCS1,CNCS2} at the Spallation Neutron Source at Oak Ridge National Laboratory (ORNL).
The data were collected on a single crystal \bro\ sample with a mass around 0.5~g, which was aligned in three different orientations: $(hk0)$, $(h0l)$ and $(0kl)$ scattering planes.
We used two fixed incident neutron energies of $E_{\mathrm{i}}=12.0$~meV ($\lambda_{\mathrm{i}}=2.61$~{\AA}) and $E_{\mathrm{i}}=3.32$~meV ($\lambda_{\mathrm{i}}=4.96$~{\AA}) resulting in a full-width-at-half-maximum (FWHM) energy resolution at the elastic position of 0.7~meV and 0.11~meV respectively.
The measurements were performed using the rotating single-crystal method.
All time-of-flight datasets were combined to produce a four-dimensional scattering-intensity function $I(\mathbf{Q},\hbar\omega)$, where $\mathbf{Q}$ is the momentum transfer and $\hbar\omega$ is the energy transfer.
For data reduction, analysis, and linear spin wave theory calculations, we used the \textsc{Mantid}~\cite{Mantid}, \textsc{Horace}~\cite{Horace} and \textsc{SpinW}~\cite{SpinW} software packages.

\section{Results and analysis}

\subsection{Bulk Properties}

\begin{figure*}[tb]
\includegraphics[width=0.98\linewidth]{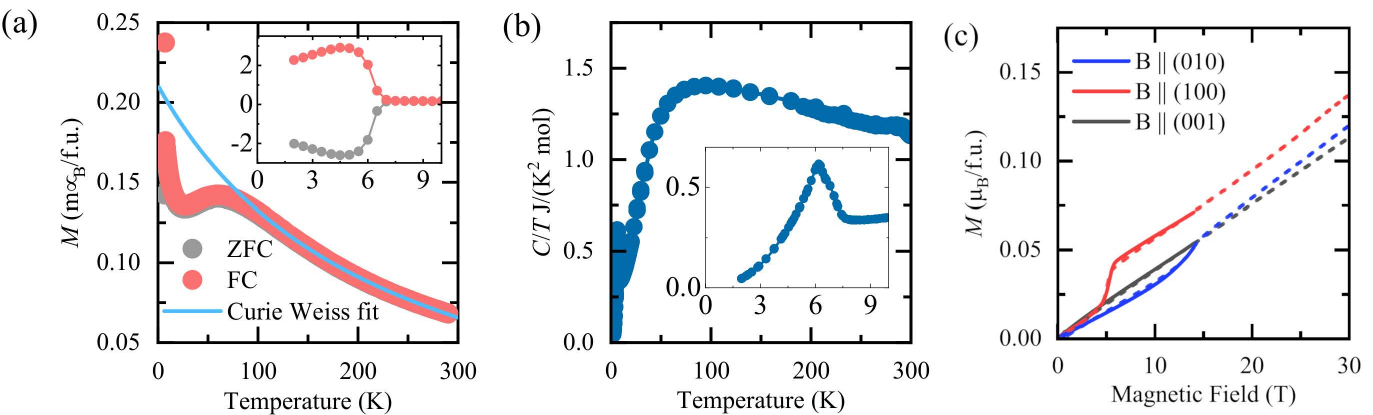}
  \caption{
  (a)~Temperature dependence of magnetization, $M(T)$, measured in an  external field of 100~Oe applied in arbitrary direction in FC and ZFC regimes. The blue line shows the fit of the high-temperature part, $T > 150$~K, with Curie-Weiss law that yields $T_{\mathrm{CW}}=-79(5)$~K.
  (b)~Temperature dependence of normalized specific heat, $C/T$, taken at zero field.
  (c)~Field dependence of magnetization measured along three principal directions at $T = 1.5$~K. The low-field data shown by solid lines were measured in DC field PPMS. The high-field data plotted by dashed lines were obtained using the pulsed-field setup. As there is no hysteresis, only the field-up sweep is shown.
  Insets at panels (a) and (b) zoom the temperature region near the phase transition.}
  \label{Fig:thermodynamics}
    \vspace{-12pt}
\end{figure*}

We start the presentation of our results with a brief overview of the thermodynamic data measured on \bro.
Figure~\ref{Fig:thermodynamics}(a) displays the temperature dependence of the magnetization, $M(T)$, measured in an external field of 100~Oe in zero- (ZFC) and field-cooled (FC) regimes.
In the insert of Fig.~\ref{Fig:thermodynamics}(a), one can see a clear transition represented by an abrupt down- (up-)turn of the magnetization in ZFC and  FC data, respectively.
The transition temperature determined using the derivative corresponds to 6.3(1) and 6.0(1)~K for the ZFC and FC curves.
The magnetization collected within 30--150~K shows a broad bump around 60~K which can be attributed to short-range low-dimensional AFM correlations.
Both low- and high-temperature features have been observed in the synthetic \bro~sample~\cite{Vilminot}.
However, contrary to the synthetic powder sample, the natural mineral does not show a magnetization anomaly at 18~K confirming that it was not intrinsic to the compound in agreement with the interpretation of Ref.~\cite{Vilminot}.
The high-temperature part of the magnetization can be fitted with the Curie-Weiss law giving a Curie-Weiss temperature of $\Theta_{\mathrm{CW}} = -79(5)$~K pointing to dominant AFM interactions in the sample.

The temperature dependence of the specific heat is shown in Fig.~\ref{Fig:thermodynamics}(b). A sharp $\lambda$-shape peak is observed at 6.2(1)~K indicating a transition into the magnetically ordered state in agreement with magnetization. Our data are in good agreement with those from Refs.~\cite{Bissengaliyeva,Vilminot}.
We further note that the magnetic ordering temperature in brochantite is comparable with that reported in some other spin-chain Cu-based minerals, such as dioptase~\cite{podlesnyak2019magnetic, prokhnenko2021high} or antelerite~\cite{kulbakov2022incommensurate, inosov2018quantum}.

The field dependence of the magnetization measured at base temperature is depicted in Fig.~\ref{Fig:thermodynamics}(c). To get an absolute magnetization value, the pulsed-field data were calibrated by the data measured at DC field using PPMS. While for the field applied along the $\bf{c^{*}}$-axis the magnetization increases linearly with the field up to 30~T, there are two anomalies observed for the field applied along the other two crystallographic directions.
For the field applied along the $\bf{a^{*}}$-direction, there is a metamagnetic-like transition at $B_C\approx$ 5~T and for $B\|\bf{b^{*}}$ a weaker jump in magnetization is detected at 13.9~T (in both cases the critical field, $B_{\rm c}$, is determined from the derivative, $dM/dB$). This suggests that the AFM easy axis lies in the $ab$-plane close to the $\bf{a^{*}}$-direction.

\subsection{{Neutron diffraction} \label{sec:neutron diffr}}

Detailed investigations of the crystal structure have been carried out by means of single-crystal neutron scattering experiments at the Institute Laue Langevin (ILL) in Grenoble (France). For this purpose, a cube-shaped crystal with a side size of 4~mm has been cut from a larger crystal. In the beginning, the reciprocal space survey at 15 and 1.5~K, i.e. above and below the magnetic ordering temperature, has been carried out using Laue diffraction at the Cyclops instrument.
% (Fig~\ref{Laue}).
Our findings at 15~K are in agreement with the literature~\cite{Merlino}. The obtained lattice parameters are $a=$13.122, $b =$ 9.838, $c =$ 6.030~\AA, $\beta =$ 103.39$^{\circ}$.
%In agreement with the literature, the observed peaks at 15~K can be indexed using SG P2$_1$/$a$ with $a=$13.122, $b =$ 9.838, $c =$ 6.030~\AA, $\beta =$ 103.39$^{\circ}$~\cite{Merlino}.
Similar to the other works we have found significant broadening and doubling of the peaks in the $a^{\ast}$-direction as a result of the disordered nature of the mineral under investigation~\cite{Merlino}.
%. These phenomena are well-known in brochantite-family minerals and attributed to the order-disorder nature of this mineral~\cite{Merlino}. In such layered materials, neighboring layers can be arranged in many different geometrically and energetically equivalent ways. As a result, either disordered or fully ordered sequences can appear. Obviously, the mineral under investigation here shows a disordered nature.

Further studies of the crystal structure have been performed at D9 hot neutron four-circle diffractometer at the ILL operated with a wavelength of $\lambda$ = 0.837~\AA. In total, 496 Bragg reflections at 2 K, 709 at 12 K, and 147 reflections at room temperature (as reference) have been recorded. In agreement with the Laue data, we found that the sample exhibits order-disorder structural transitions and twinning, as evidenced by the observation of peak broadening and doubling, respectively~\cite{Merlino}.
The recorded data were used for the refinement of the crystal structure. Here, an averaged structure has been treated, i.e. the intensity of both twins was integrated and the disorder has not been considered. Data analysis proves that the space group P$2_1/a$ can describe the structure measured at 12~K.

Turning to the magnetic order, the Laue data show no difference between the patterns measured at 15 and 2~K. The same observation is valid for the D9 data between 12 and 2~K. This means no additional (structural and/or magnetic) reflections appear below the magnetic transition at 6.3 K. This points out that the magnetic contribution is quite small as expected for the Cu$^{2+}$ ion, and located on the top of the nuclear peaks.

Further investigations of the magnetic structure have been carried out using D10 neutron four-circle diffractometer at the ILL which is optimized for studying magnetic structures. This instrument provides relatively high flux at 2.36~\AA\ and low intrinsic background due to the energy analysis option. Here, 203 Bragg reflections at 2~K and 59 at 15~K have been recorded. The first observation is that at 15~K some small forbidden reflections in the space group P2$_1$/$a$ are detected as, for example, (100), (300), and (-102).
Taking into account these reflections, the true symmetry of \bro\ should be lower than anticipated and could be described by space group P-1. This is a completely new finding which implies that the crystal structure of brochantite should be revised for either correction or looking for a symmetry lowering between RT and 15 K~\cite{Merlino}.
Our data-set focused on a low momentum transfer part of the reciprocal space leaves this problem out of the scope of this work.
%does not contain enough reflections to deal with this problem quantitatively. }
However, because the deviations from the parent P2$_1$/$a$ space group are small and for the sake of simplicity we will perform the representation analysis of possible magnetic structures using P2$_1$/$a$ space group.

\begin{figure}[tb]
\includegraphics[width=0.6\linewidth]{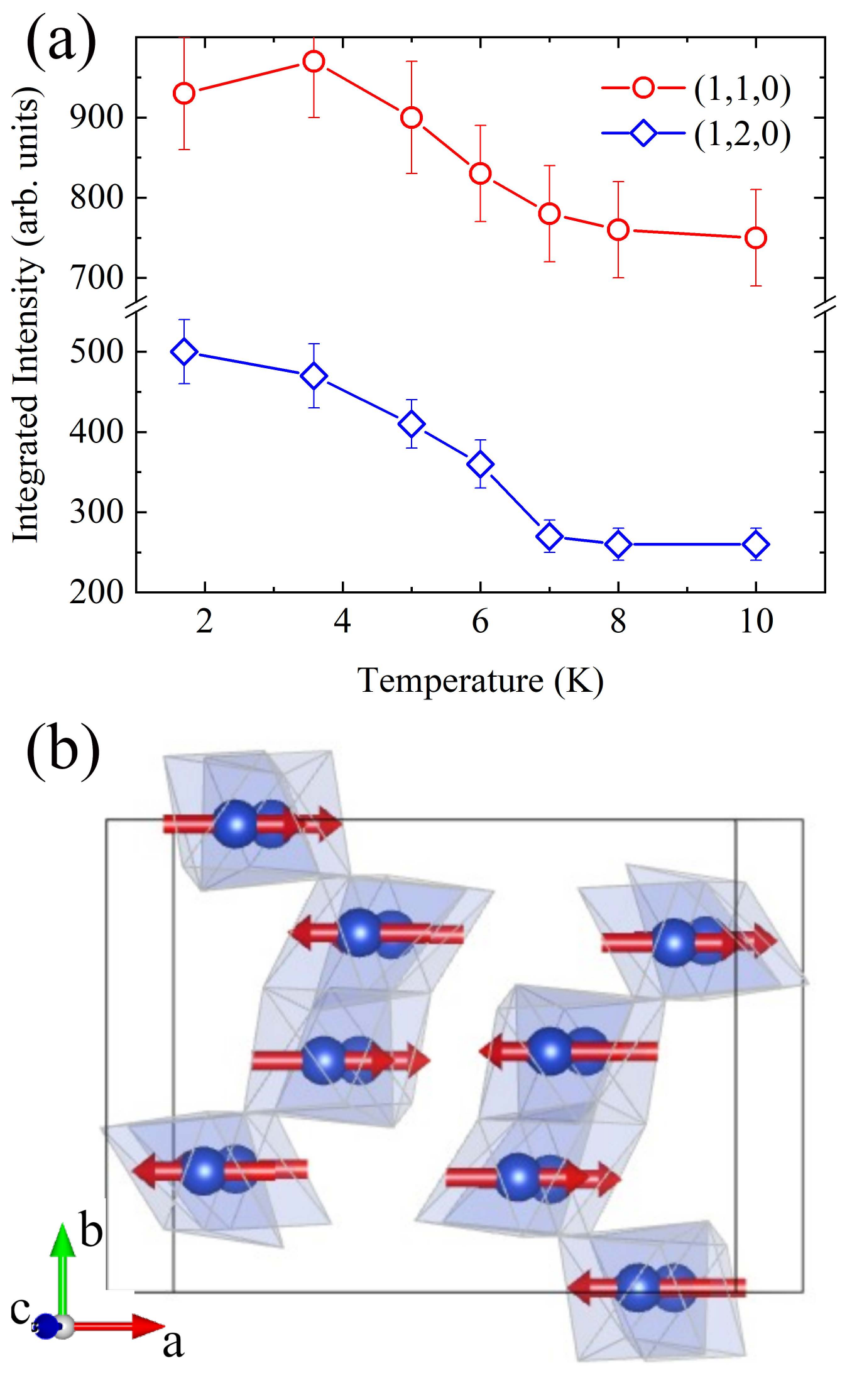}
  \caption{~(a) Integrated intensity of the (1,1,0) and (1,2,0) reflections, showing the magnetic contribution as a function of temperature. (b) Magnetic structure of \bro. Only Cu ions are shown. View along the $\bf{c^*}$-axis.
  }
  \label{MagStr}
  \vspace{-12pt}
\end{figure}

At 2~K, in the magnetically ordered phase, the increase of the intensity due to magnetism is noticeable only on the (110) and (120) reflections, see Fig.~\ref{MagStr}(a). The change in intensity of the remaining reflections is within the error bars. However, the temperature dependence of the two relevant reflections shows clearly the magnetic transition at about 7~K in agreement with the bulk data. This allows us to conclude that the magnetic propagation vector is $\mathbf{Q}_{\rm m}$=(0,0,0). This is in agreement with the results obtained by neutron powder diffraction on the synthetic sample~\cite{Vilminot}. Unfortunately, the presence of 16 independent magnetic ions renders the proper refinement of the magnetic structure, in this case, impossible. However, by using all the available bulk and neutron data and testing a suite of reasonable quantitative magnetic structures, a model of the magnetic structure can be proposed.

\begin{table}
\caption{ \label{Irreps} Irreducible representation of the little group $\bf{G_k}$ containing all the symmetry elements which leave $\mathbf{Q}_{\rm m}$=(0,0,0) invariant.}
\begin{ruledtabular}
    \begin{tabular}{cccccc}
                    & 1 & 2$_1$ & $\bar{1}$ & $a$ & Magnetic SG  \\
           \hline
           $\Gamma_1$  & 1  & 1  &  1 &  1 &  P2$_1$/$a$ \\
           $\Gamma_2$  & 1  & 1  & -1 & -1 &  P2$_1$/$a^\prime$ \\
           $\Gamma_3$  & 1  & -1 &  1 & -1 &  P2$_1^\prime$/$a^\prime$ \\
           $\Gamma_4$  & 1  & -1 & -1 &  1 &  P2$_1^\prime$/$a$ \\
    \end{tabular}
\end{ruledtabular}
\end{table}

\begin{figure*}[tb]
\center{\includegraphics[width=0.9\linewidth]{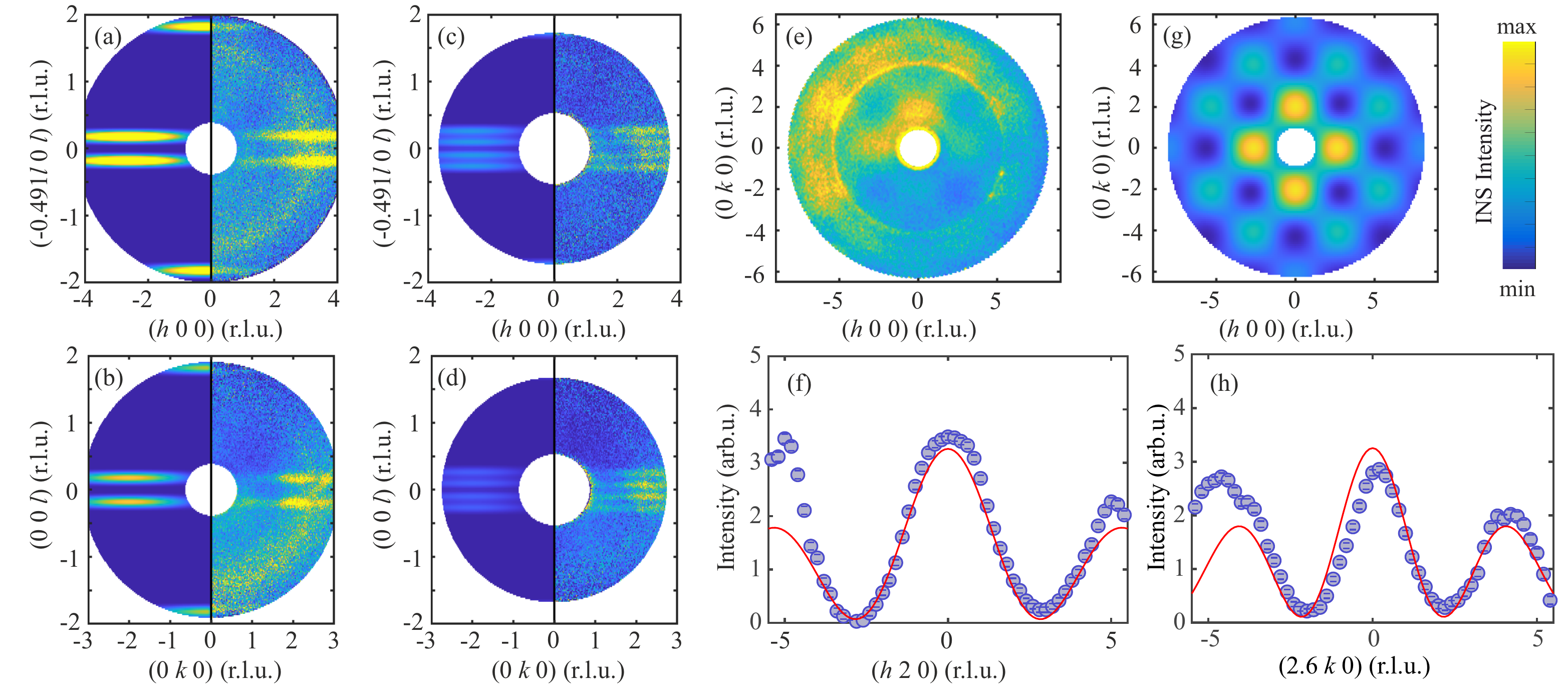}}
  \caption{~Constant-energy slices showing one-dimensional magnetic excitations.
  (a-d)~Constant-energy slices within two scattering planes measured on the CNCS spectrometer with $E_{\rm i} = 3.32$~meV. The spectra in panels (a,b) were integrated within $E = [1.2$--$1.5]$~meV and in panels (c,d) within $E = [2.2$--$2.5]$~meV. The left and right portions of each panel show the results of LSWT calculations, and the experimental data respectively.
  (e,g)~Constant-energy slice at $E = [2.5$--$4.5]$~meV measured on the CNCS spectrometer with $E_{\rm i}= 12$~meV (e) and calculated using LSWT (g). Orthogonal wavevector integration window $l = [$-$0.3$--$0.3]$~r.l.u. The asymmetric distribution of the INS intensity in (e) is due to the anisotropic shape and absorption of the sample.
  (f,h)~One-dimensional cuts along $(h~2~0)$ and $(2.6~k~0)$ directions of the reciprocal space. A linear background was subtracted from the data. Red lines show LSWT calculations for the zig-zag chain. Wave vectors are indexed in reciprocal lattice units (r.l.u.).
  }
  \label{fig_str_factor}
  \vspace{-12pt}
\end{figure*}

Table~\ref{Irreps} contains the irreducible representations (irreps) which can be used to define possible magnetic structures with  $\mathbf{Q}_{\rm m}$=(0,0,0). According to Vilminot~{\it{et~al.}}~\cite{Vilminot}, the ground state magnetic structure can be described by almost linear FM Cu-chains running along the $c$-axis.
In turn, nearest-neighbor chains couple into an AFM zig-zag double chain, which is in agreement with irrep $\Gamma_3$ (magnetic SG P2$_1^\prime$/$a^\prime$) applied to all four Cu-sites.  Interestingly, to describe their NPD data the authors assumed different Cu-moment sizes within the chains. However, this would lead to a substantial uncompensated FM contribution which should be detected by magnetization measurements.
As the latter is not the case, the models to be considered should only be the AFM ones. Nevertheless, we have tested this model with equal moment sizes and found that it does not describe our experimental data. Our data indeed suggest that the moments are arranged FM in the chains along the $c$-axis and these chains are coupled AFM. However, contrary to the Vilminot's model, the inversion symmetry should be primed (i.e. combined with time-reversal), leaving two possible magnetic space groups P2$_1$/$a^\prime$ and P2$_1^\prime$/$a$ corresponding to $\Gamma_2$ and $\Gamma_4$, respectively. Given the very limited number of reflections with magnetic contribution and cross-correlation between the moment size and its orientation, the unique determination of the magnetic structure is impossible.

Taking into account the magnetization data, one would assume the moment direction to be close to the $a$-axis, resulting in the magnetic SG P2$_1^\prime$/$a$  ($\Gamma_4$). Restricting ourselves to a collinear magnetic structure, one gets the one shown in Fig.~3(b) with an estimated magnetic moment of about 0.30$\pm$0.15~$\mu_B$.
To obtain this value, the size of the magnetic moment was adjusted such that the respective growth of the nuclear (110) and (120) reflections corresponds to the one observed experimentally.
In addition, the symmetry analysis shows that the antisymmetric DM components of the nearest (NN) and next-nearest neighbor (NNN) exchange matrix are allowed for Cu-Cu interaction.
Therefore, we would rather expect an equal or very similar moment size with possible spin-spiral correlation given by the uniform DM interactions.

\subsection{{Inelastic neutron scattering}\label{sec:ins}}

While the neutron diffraction and thermodynamic measurements in brochantite provide evidence for a spin chain system with a possible DM interaction, the details of the exchange interactions have to be determined by INS.
The constant energy slices in $(hk0)$,$ (h0l)$ and $(0kl)$ scattering planes obtained using INS measurements on a single crystal of \bro\ are summarized in Fig.~\ref{fig_str_factor}.
Two different incident neutron energies were used, $E_{\rm i} = 12.0$~meV and 3.32~meV, to combine a large accessible energy-transfer range at higher energies with an improved resolution at lower energies.
As shown in Fig.~\ref{fig_str_factor}(a-d), the dispersion along the $l$ direction is strong, while little dispersion is observed in other, $h$ and $k$ directions.
Note, that the weak dispersion along the $h$ direction can be the result of structural disorder.
The considerable dispersion along $l$ and almost flat branches along both $h$ and $k$ directions indicate strong 1D character and weak interchain coupling.
Another distinct feature of the measured magnetic excitations is the checkerboard-shaped pattern in $(hk0)$ constant-energy slices of $S(\mathbf{Q},\hbar\omega)$ [Fig.~\ref{fig_str_factor}(e-h)].
A similar phenomenon was previously observed in the quasi-one-dimensional compounds with 1D zig-zag chain, CoNb$_2$O$_6$ \cite{Cabrera} and YbFeO$_3$ \cite{Nikitin}.
The chain buckling leads to the BZ folding due to an increase of the magnetic unit cell that, in turn, results in the checkerboard pattern of the structure factor of the magnetic excitations.
In \bro, pairs of Cu atoms connected to each other, Cu1-Cu2 and Cu3-Cu4, give rise to nearly linear chains coupled by $J_2$, which in turn couple into the zig-zag double chain via $J_1$, as shown in Fig.~\ref{spin-chain}(c).
The experimentally observed checkerboard pattern suggests that $J_1 \gg J_2$, i.e. is attributed to the formation of the {\it{zig-zag}} magnetic chains along the $c$ direction.

The experimental spectra along $l$ direction consist of strong magnon-like modes merging into an extensive continuum.
The magnetic branch could be traced up to an energy transfer of $\sim$7~meV due to strongly reduced intensity at higher energies.
The magnons' minima are located at $l = 0 \pm 0.19$~r.l.u. suggesting spiral spin-chain dynamics in spite of experimentally observed commensurate long-range AFM order.
A finite spectral gap $\Delta = 1.0(1)$~meV was observed in zero-field spectra at temperatures below $T_{\mathrm{N}}$, which implies the presence of anisotropic couplings.

In low-dimensional systems with weak magnetic order such as weakly-coupled spin chains, the low-energy excitations can be considered as conventional magnons.
The high-energy sector in turn is dominated by deconfined spinons and there is a crossover regime between the two limits~\cite{lake2005quantum}.
As shown above, the low-energy part of the INS spectrum is dominated by sharp spin-wave excitations.
In contrast, the high-energy spectrum of brochantite demonstrates broad continuous excitations at energies above 5~meV, which can be clearly seen in constant-energy cuts, Fig.~\ref{ins}(c), and we interpret as the presence of broad continua as a possible manifestation of deconfined spinons~\cite{Muller, Tennant}.

\subsection{{LSWT simulations and discussion}\label{sec:lswt}}

In order to describe the main properties of the observed spectra we consider a $S$=1/2 XXZ AFM Heisenberg chain supplemented by a symmetric anisotropic exchange interaction and an antisymmetric DM interaction.
The Hamiltonian is given by
\begin{eqnarray}
\mathcal{H} &=& J_1\sum_i (S^x_iS^x_{i+1} +  S^y_iS^y_{i+1} + \delta S^z_iS^z_{i+1}) \nonumber \\
&+&  J_2\sum_i \mathbf{S}_i \mathbf{S}_{i+2} - D\sum_i (S^x_iS^y_{i+1} -  S^y_iS^x_{i+1}),
\label{XXZ}
\end{eqnarray}
where $J_1$ and $J_2$ are the NN and NNN exchange couplings, $\delta>1$ is the easy-axis anisotropy, and $D$ is the antisymmetric DM interaction pointing along the easy axis.

\begin{figure}[tb]
\center{\includegraphics[width=1\linewidth]{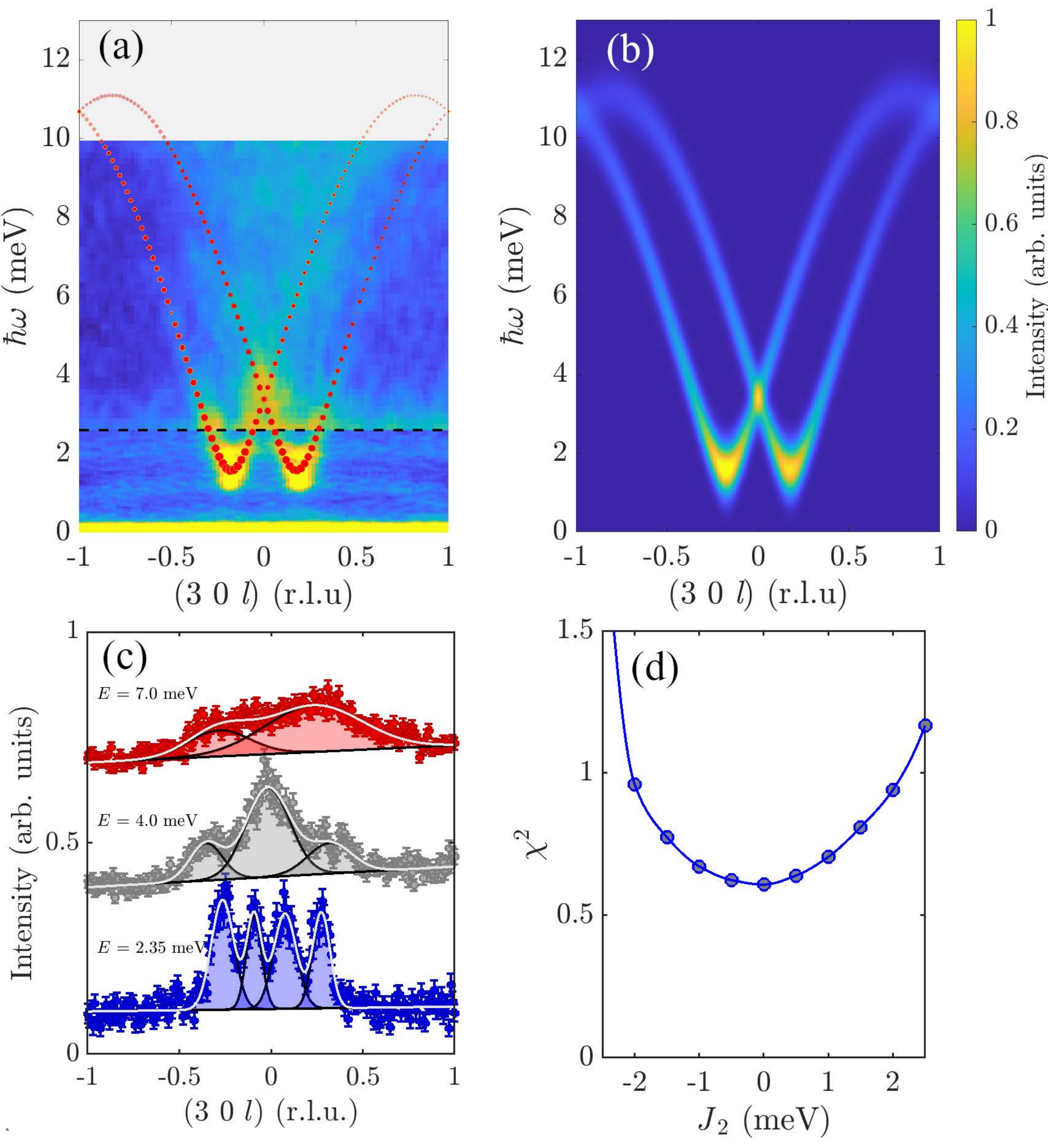}}
  \caption{~(a) Spin-wave dispersion of \bro\ in AFM state at 1.7~K measured at CNCS along $l$ direction. The spectrum's lower part below the black dashed line was collected with $E_{\mathrm{i}} = 3.32$~meV and the upper part with $E_{\mathrm{i}} = 12$~meV.
  The data are integrated by $\pm0.5$~r.l.u. in two orthogonal directions. Red points show the dispersions fitted by LSWT and their sizes represent the calculated intensities. (b) The INS intensity (the magnon spectral weight) along $l$ direction calculated by LSWT using the best-fit parameters of Hamiltonian (\ref{XXZ}), see text. (c) One-dimensional constant-energy cuts at different energy transfers. The data were obtained by integration of the INS signal within $h = 3 \pm 0.5$, $k = 0 \pm 0.5$ r.l.u. at $E = 2.35 \pm 0.15$, $4 \pm 0.5$ and $7 \pm 0.5$~meV. The data are shifted vertically for clarity. The data were fitted using sum of several Gaussian functions and linear background. Grey solid lines demonstrate total fit to the data. Black lines along with shaded areas demonstrate contributions of individual Gaussian peaks to the total fit.
  (d) The reduced $\chi^2$ value from the best fit for different $J_2$, see text.
  }
  \label{ins}
\end{figure}

We start our analysis with the case $D=0$.
This is one of the most investigated models that deals with the symmetric frustrating intrachain couplings, where the NN exchange $J_1$ competes with the AFM NNN $J_2 > 0$ exchange~\cite{Majumdar1969, Haldane, Villain, Yoshimori, Chubukov, Hikihara, Sudan, Mikeska2004, Okamoto, Eggert, Tonegawa,White}.
The model develops a variety of exotic phases like vector-chiral, spin-nematic, or higher-order polar phases~\cite{Chubukov,Hikihara,Sudan}.
The ground state of the quantum spin chain undergoes a phase transition from a phase with gapless excitation for $J_2 \lesssim 0.24J_1$ to a gapped phase for the larger $J_2$~\cite{Mikeska2004,Majumdar1969,Haldane,Okamoto,Eggert,White}.
Incommensurate spiral correlations are present for $J_2 > 0.5J_1$~\cite{Tonegawa, White}, which exhibit a qualitatively similar excitation spectrum, such as was found in the double-helix material FeP~\cite{sukhanov2022frustration}.
In brochantite, considering only symmetric exchange interaction $J_1 \gg J_2$, we expect a standard N\'{e}el ground state stabilized by the easy-axis magnetic anisotropy.
In this approximation, the spin-spiral correlations are implausible.

In case $D \neq 0$, the helical spin dynamics can arise due to competition between the symmetric and antisymmetric interactions.
A uniform DM vector, collinear with the easy axis and with a magnitude less than some critical value, favors spin-spiral correlations in the otherwise commensurately ordered phase~\cite{Santos1, Santos2}.
The double-degenerate spin-wave modes with opposite angular momenta are shifted symmetrically, so their excitation energy minima are centered at $l = 0 \pm q$, where $q$ is the wave vector of the spiral.

It is well known that linear spin-wave theory (LSWT) provides only limited qualitative agreement with the measured
spin dynamics of low-dimensional quantum magnets.
However, LSWT can qualitatively capture the dispersion and magnon bandwidth of the ordered magnets throughout the Brillouin zone.
Besides, having low computational cost, LSWT calculations can be used to check various scenarios, including systems with multiple couplings.
We use LSWT to determine the parameters of Hamiltonian \eqref{XXZ}, calculating a neutron scattering cross section for the spin-ladder model, Fig.~\ref{spin-chain}(c).

Since we do not observe an upper boundary of the excitation, we use the experimental low-energy spectrum $\hbar\omega < 7$~meV, which exhibits a number of important features, such as branch stiffness, the anisotropy gap, and the wave vector of the spiral, that we used to determine the parameters of Hamiltonian~(\ref{XXZ}).
We quantify the energy of the sharp magnon mode at 23 different points in reciprocal space $0 < l < 1$ r.l.u. and use this experimental data set to fit the exchange couplings.
The INS spectrum was calculated using an equation for the nonpolarized neutron scattering cross section as implemented in the SpinW \cite{SpinW, Squires_2012}.
The chi-square fit of the experimentally observed and calculated INS spectra yields $J_1 = 10.6(2)$~meV, $J_2 = 0(2)$~meV, $\delta = 1.05(1)$, and $D = 3.0(2)$~meV.
The uncertainties were estimated by comparing the exchange interactions obtained from fits with different initial parameters.
One can see an excellent agreement between the calculated and experimental spectrum (Fig.~\ref{ins}), including the general shape of the dispersion curve and the spectral intensity over reciprocal space.

Surprisingly, the best agreement between the simulations and the data was achieved for the NNN coupling $J_2 = 0$~meV and rather large $D = 0.28J_1$.
This suggests that a single zig-zag chain model can well describe the experimentally observed spin-spiral dynamics.
However, within the precision of our measurements, we cannot exclude the presence of a small NNN term.
Specifically, fixing the $J_2$ at some non-zero value ($|J_2|<2.5$) and adjusting $J_1$ and $D$, we still can find a reasonably good solution with the reduced $\chi^2$ value slightly larger ($\sim$1-2\%) when compared to the best fit, see Fig~\ref{ins}(d).
Yet, any increase of the $J_2$ consistently worsens the fit.
Clear visual disagreement between the calculated and observed spectra was found for $|J_2| > 2.5$~meV.

It is important to mention that the magnon excitations observed in both the $h$ and $k$ directions exhibit weak dispersion.
This indicates that interchain exchange interactions are present and stabilize long-range magnetic order at $T_{\rm N} = 6.2$~K, as expected for weakly-coupled spin-chain system~\cite{schulz1996dynamics}.
The interchain exchange can cause a splitting of the degenerate modes, even in the absence of an applied magnetic field.
However, this splitting could not be unambiguously resolved in our experiment due to its small magnitude.

\section{Conclusions}

Summarizing, we have investigated the magnetic properties of the natural mineral brochantite using magnetic neutron diffraction, thermodynamic measurements, and inelastic neutron scattering.
The key feature of the spin subsystem of brochantite is the uniform DM chiral coupling that leads to spin-spiral dynamics.
Nonetheless, as the magnitude of antisymmetric interaction is below some critical value, the ground state of \bro\ is commensurate AFM, stabilized by easy-axis exchange anisotropy.

The LSWT calculation shows good agreement with the observed low-energy magnon excitation, enabling us to quantify the proposed model's symmetric and antisymmetric exchange interactions.
While LSWT correctly predicts the ground state and single magnon excitation, it fails to reproduce the observed high-energy continuum.
The quantum effects due to low dimensionality and low spin value must be considered to describe the full experimental spectra.
We note that the quantum effects also cause the bandwidth renormalization and the exchange interaction deduced by LSWT should be divided by $\pi/2$ factor to obtain the true interaction strength~\cite{des1962spin} in the case of a simple Heisenberg model.
DM interaction qualitatively modifies the spinon behavior and produces new phenomena such as backscattering interaction between the spinons~\cite{povarov2022electron} and unusual spin excitations~\cite{wang2022hydrodynamics}.
This calls for further experimental and theoretical studies including polarized INS in fields and density matrix renormalization group calculations; this work is in progress.

\section*{Acknowledgments}
The authors thank Daniel Pajerowski for the helpful discussions, Jong Keum for assistance with x-ray Laue measurements. Work at ORNL was supported by the U.S. Department of Energy (DOE), Office of Science, Basic Energy Sciences, Materials Science and Engineering Division. X-ray Laue alignment and magnetization measurements were conducted at the Center for Nanophase Materials Sciences (CNMS) at ORNL, which is a DOE Office of Science User Facility. This research used resources at the Spallation Neutron Source, a DOE Office of Science User Facility operated by ORNL. Work by LMA was supported by the U.S. Department of Energy, Office of Science, Office of Basic Energy Sciences, Chemical Sciences, Geosciences, and Biosciences Division. SEN acknowledges European Union Horizon 2020 research and innovation program for Marie Sklodowska-Curie Grant No. 884104 for financial support. OP acknowledges the GIMRT Program of the Institute for Materials Research, Tohoku University (Proposal No. 20K0507).

\bibliography{bibliography}

\end{document}